\documentclass[aps,prl,twocolumn,groupedaddress]{revtex4-2}
\usepackage{amsmath,amssymb,amsthm,mathrsfs,amsfonts,dsfont} 
\usepackage{bm,booktabs,enumitem,placeins}
\usepackage{graphicx}
\usepackage{subfigure}
\usepackage{amsbsy}
\usepackage{bm}
\usepackage{orcidlink}
\usepackage{xcolor}

\usepackage{algorithm}
\usepackage{algpseudocode}
\usepackage{algorithmicx}

\newcommand{\avg}[1]{\langle #1\rangle}

\begin{document}

\title{Speed-driven plasticity tunes chaotic fluctuations: dynamical mean-field theory and generalized Schr\"odinger equation}
\author{Xinrui Wen$^{1}$}
\author{Haiping Huang\,\orcidlink{0000-0001-8757-4733}$^{1,2}$}
\email{huanghp7@mail.sysu.edu.cn}
\affiliation{$^{1}$PMI Lab, School of Physics, Sun Yat-sen University, Guangzhou 510275, People's Republic of China}
\affiliation{$^{2}$Guangdong Provincial Key Laboratory of Magnetoelectric Physics and Devices, Sun Yat-sen University, Guangzhou 510275, People's Republic of China}
\date{\today}

\begin{abstract}
	Synapses in neural circuits can strongly interact with rapidly changing neural activity. Recent theoretical and computational studies revealed that speed-driven synaptic plasticity plays an important role in biological dendritic predictive learning and memorization-to-generalization in recurrent generative models. Here, we characterize this joint system of both neurons and synapses by developing dynamical mean-field theory and a generalized Schr\"odinger equation. Three insights from the theory are provided. First, speed-drive plasticity does not alter the chaotic transition. Second, negative plasticity acts as a brake on the neural activity. Third, positive plasticity produces a preferred frequency. All emergent behaviors are predicted by our theory.
\end{abstract}

 \maketitle

\textit{Introduction.}---
Recurrent neural networks are candidate theoretical and computational models of highly non-linear, multi-scaled brain dynamics~\cite{Vyas-2020,Ostojic-2024,Huang-2024,DWK-2025}. The network structure evolves during state-based recurrent computations~\cite{Abbott-2009,Maass0-2009}. This is akin to spin-coupling interaction models studied in physics~\cite{JPA-1993,JPA-1994}, which are much more complex than a pure spin system. A recent generalization to neuronal systems considers Hebbian learning in classic random neural networks, where the baseline coupling is independently Gaussian distributed. The symmetric Hebbian learning was found to be able to stabilize chaos~\cite{Clark-2024,Jstat-2023,Du-2024}. However, freezing chaos can also be achieved by Onsager dynamics feedback by introducing gradient descent dynamics on the kinetic energy surface~\cite{Qiu-2024,Huang-2025,Zhang-2026}. This recent kinetic energy perspective yields asymmetric speed-driven synaptic plasticity, in particular in the generative model, namely the neural Langevin machine, and in memory storage via this type of asymmetric learning~\cite{Yu-2025}. Moreover, the speed-driven plasticity has a biological correspondence in dendritic predictive learning~\cite{Senn-2014,Senn-2024}, such as reducing the prediction error of somatic spiking. How the chaotic fluctuation (even the transition point) is modified in the neuron-synapse coupled system under speed-driven asymmetric plasticity remains elusive. Revealing the underlying mechanism may help to generate biologically testable hypotheses and to improve generative model performance and even understand memory capability. 

Here, we develop a generalized dynamical mean-field theory for this neuronal-synaptic coupled system, whose unlearning counterpart reduces to the classic random neural networks~\cite{Chaos-1988}. To identify whether the chaotic transition is shifted, we develop a generalized Schr\"odinger equation whose ground-state energy determines the separation speed of two close-by neural trajectories subject to learning. We surprisingly found that the learning does not change the transition point, in contrast to recent related studies~\cite{Clark-2024,Du-2024}. Moreover, the plasticity strength, including its sign, tunes the chaotic fluctuations. The maximal Lyapunov exponent displays non-monotonic behavior with the strength. Negative plasticity slows down the fluctuation, while positive plasticity causes single neurons to oscillate in their own frequencies. All the observations are predicted by our theory, further supported by simulations on finite-sized networks.

\textit{Model.}---
We consider a random-network model consisting of $N$ interacting units, whose dynamical state is described by $\bm{x}(t)\in\mathbb{R}^N$. The synaptic coupling between neurons is subject to speed-driven plasticity. The network evolves according to the following coupled synaptic-neuronal dynamics:
\begin{align}
    \dot{\bm{x}}(t)
    &= -\bm{x}(t)
    + \bm{J}(t)\phi(\bm{x}(t)),
    \label{eq:network_dynamics}\\
    \bm{J}(t)
    &= \bm{J}_0+\bm{J}_1(t),
    \label{eq:coupling_decomposition}\\
    \tau_s\dot{\bm{J}}_1(t)
    &= -\bm{J}_1(t)
    +\frac{\lambda}{N}
    \dot{\bm{x}}(t)\phi(\bm{x}(t))^{T},
    \label{eq:plasticity_dynamics}
\end{align}
where $\phi$ is a nonlinear activation function applied componentwise, and $\bm{J}(t)\in\mathbb{R}^{N\times N}$ denotes the time-dependent connectivity matrix.
Time is measured in units of the intrinsic neuronal relaxation time,
so that $\tau_s$ is the ratio of the plasticity and neuronal relaxation
timescales. Unless stated otherwise, we use $\phi(x)=\tanh(x)$.
All the following analyses use $\tau_s=2.5$ unless stated otherwise. $\tau_s>1$ is consistent with the fact that synaptic plasticity is commonly slower than neuronal dynamics.

The connectivity matrix is decomposed into two components: a quenched random matrix baseline $\bm{J}_0$ and a dynamically evolving matrix $\bm{J}_1(t)$. The elements of $\bm{J}_0$ are independently drawn from a Gaussian distribution,
\begin{equation}
    (J_0)_{ij}
    \overset{\mathrm{i.i.d.}}{\sim}
    \mathcal{N}\left(0,\frac{g^2}{N}\right),
    \label{eq:random_connectivity}
\end{equation}
where $g$ controls the strength of the random recurrent interactions. The evolution of $\bm{J}_1(t)$ is governed by Eq.~\eqref{eq:plasticity_dynamics}, which introduces a speed-dependent plasticity mechanism. The parameter $\lambda$ determines the strength of the plasticity, while $\tau_s$ characterizes its relaxation timescale. In particular, the plasticity term couples the instantaneous speed of the network state, $\dot{\bm{x}}(t)$, to the neuronal activity $\phi(\bm{x}(t))$. This is inspired by recent neural Langevin machine~\cite{Yu-2025}, and moreover biological dendritic predictive learning~\cite{Senn-2014,Senn-2024}.

The initial conditions are specified as $\bm{J}_1(0)=\bm{0}$, and $x_i(0)\overset{\mathrm{i.i.d.}}{\sim} \mathcal{N}(0,\sigma^2)$, $\forall i$, 
where $\sigma^2$ denotes the variance of the initial neuronal states. The initial states and the quenched random connectivity are sampled independently.

\textit{Dynamical mean-field theory.}---
We first insert the formal solution of Eq.~\ref{eq:plasticity_dynamics} into the synaptic integration term $\bm{J}(t)\phi(\bm{x}(t))$, and then take the large-$N$ limit, obtaining the following single-site effective mean-field equation:
\begin{equation}
    \dot{x}(t)+x(t)
    =\eta(t)
    +\frac{\lambda}{\tau_s}
    \int_{-\infty}^{t}
    e^{-\frac{t-s}{\tau_s}}
    C_\phi(t-s)\dot{x}(s)\,ds,
    \label{eq:effective_dynamics}
\end{equation}
where $x(t)\in\mathbb{R}$ represents the state of a typical neuron in the network. The effective driving field $\eta(t)$ is a stationary Gaussian process with zero mean and covariance:
\begin{align}
    \langle\eta(t)\rangle_{\eta}&=0,
    \label{eq:noise_mean}\\
    \langle\eta(t)\eta(t+\tau)\rangle_{\eta}
    &=g^2 C_\phi(\tau).
    \label{eq:noise_covariance}
\end{align}
Here, $\langle\cdot\rangle_{\eta}$ denotes an average over realizations of the effective Gaussian process.
The function $C_\phi(\tau)\equiv  \left\langle \phi(x(t))\phi(x(t+\tau))\right\rangle_{\eta}$ is defined as the stationary autocorrelation function of the nonlinear neuronal activity.
$C_\phi=\mathcal{G}[C_x]$, where $\mathcal{G}$ is the Gaussian map: $C_\phi(t)=\mathcal{G}\big(C_x(t);C_x(0)\big)=\int\!Dz\Big[\int\!Dy\,\phi\big(\sqrt{C_x(0)-C_x(t)}\,y+\sqrt{C_x(t)}\,z\big)\Big]^2$.
A path-integral approach to derive the above mean-field dynamics equation is presented in SM~\cite{SM}, which is also named dynamical mean-field theory (DMFT)~\cite{Zou-2024,MSR-1973,Sommers-1987,Jans-1976,Domin-1978}. But here we call it the generalized DMFT, particularly for learning in recurrent networks.

In the stationary limit, the key physical quantity such as the autocorrelation becomes time-translation-invariant. We then define the current autocorrelation $C_x(\tau)=\avg{x(t)x(t+\tau)}_{\eta}$, whose Fourier transform is named $S_x(\omega)$, while the activity autocorrelation $C_\phi(\tau)$ has its Fourier transform $S_{\phi}(\omega)$. We also define the memory kernel $K(u)=\frac{\lambda}{\tau_s}e^{-u/\tau_s}C_\phi(u)$.
 We solve the time-dependent DMFT equation in the frequency domain as follows:
\begin{equation}
S_x(\omega)=\frac{g^2S_\phi(\omega)}{|\ell(i\omega)|^2}=\frac{g^2\,S_\phi(\omega)}{\big|1+i\omega\,[1-\tilde K(i\omega)]\big|^2}, \label{eq:closure}
\end{equation}
where $\ell(z)=1+z[1-\tilde{K}(z)]$, $\tilde K(z)=\int_0^\infty K(u)e^{-zu}du$. At $\lambda=0$, $|\ell|^2=1+\omega^2$ and Eq.~\eqref{eq:closure} reduces to the classical Sompolinsky-Crisanti-Sommers (SCS) equation \textit{without}learning~\cite{Chaos-1988}.

\textit{Stability of the trivial fixed point.}---
We investigate the local stability of the trivial fixed point under the assumption $\phi(0)=0$. In this case, the coupled dynamics admits a fixed point at $( \bm{x}^{*},\bm{J}_1^{*})=(0,0)$
where the connectivity matrix reduces to $\bm{J}^{*}=\bm{J}_0$.
The local stability of this fixed point can be determined by linearizing the coupled dynamics with respect to small perturbations in $\bm{x}$ and $\bm{J}_1$. In the thermodynamic limit $N\to\infty$, the critical coupling strength is given by $g_c=\frac{1}{\phi'(0)}$~\cite{SM}. Hence,
The trivial fixed point is locally asymptotically stable for $g<g_c$ and becomes unstable for $g>g_c$. Notably, the stability threshold is independent of the plasticity strength $\lambda$ and the relaxation timescale $\tau_s$.

\textit{Maximum Lyapunov exponent and generalized Schr\"odinger equation.}---
  When the trivial fixed point becomes unstable, the system may enter chaos, causing exponential separation of two nearby trajectories. Therefore, we next determine the asymptotic maximum Lyapunov exponent. The positive exponent implies chaos. We first take two copies $a=1,2$ of the full system $(\bm x^a,J^a)$ with the same $J^0$ and different initial conditions. Averaging the path integral representation over $J^0$ produces the cross term controlled by $C^{12}=\frac1N\sum_j\phi(x^1_j)\phi(x^2_j)$ (see a similar treatment in our previous work~\cite{Yu-2025b}). Hence, the pair single-site problem is given by:
  \begin{equation}
\mathcal L_t\,x^a=\eta^a,\quad \mathcal L_t f(t)=f(t)+\dot f(t)-\int_0^\infty K(u)\dot f(t-u)du,
\end{equation}
where $\langle\eta^a(t)\eta^b(s)\rangle=g^2C_\phi^{ab}(t,s)$. Both replicas are on the same stationary attractor, so $C^{11}_\phi=C^{22}_\phi=C_\phi$ and $K$ is the stationary memory kernel.

The two replicas are coupled only through $C^{12}$, via 
\begin{equation}
\mathcal L_t\mathcal L_s\,C_x^{12}(t,s)=g^2\,C^{12}_\phi(t,s). \label{eq:cross}
\end{equation}
According to replica symmetry, $C_x^{12}(t,s)=C_x(t-s)-\epsilon(t,s)$, where $\epsilon(t,s)$ is a small deficit, so the mean squared distance is $d(t)=\frac1N|\bm x^1-\bm x^2|^2=2\epsilon(t,t)$. Price's theorem gives $\partial C^{12}_\phi/\partial C^{12}_x=\langle\phi'(x^1)\phi'(x^2)\rangle\to C_{\phi'}(t-s)$. To first order in $\epsilon$,
\begin{equation}
\mathcal L_t\mathcal L_s\,\epsilon(t,s)=g^2C_{\phi'}(t-s)\,\epsilon(t,s). \label{eq:lin}
\end{equation}
Note that the kernel $K$ itself is not perturbed, as $C^{aa}$ is exactly stationary. 
We next insert the separation ansatz $\epsilon(t,s)=e^{\Lambda(t+s)/2}\psi(t-s)$~\cite{Huang-2022} and use $\mathcal L_te^{zt}=\ell(z)e^{zt}$. We then get the generalized Schr\"odinger equation:
\begin{equation}
\Big[\big|\ell\big(\tfrac{\Lambda}{2}+i\hat p\big)\big|^2-g^2C_{\phi'}(t)\Big]\psi=E_0(\Lambda)\psi, \label{eq:lyap_main}
\end{equation}
where $\hat{p}=-i\partial_t$, $C_{\phi'}(t)=\langle\phi'(x(t))\phi'(x(0))\rangle$, and since $d\propto e^{\Lambda t}=e^{2\Lambda_{\rm max}t}$, the maximum Lyapunov exponent is given by $\Lambda_{\max}=\tfrac12\Lambda^*$, where $\Lambda^*$ is the largest root of the ground-state eigenvalue $E_0(\Lambda^*)=0$. We leave technical details to~\cite{SM}.
At $\lambda=0$, $\ell(z)=1+z$, and the eigenvalue problem becomes
\begin{equation}
\big[-\partial_t^2-g^2C_{\phi'}(t)\big]\psi=-\big(1+\tfrac\Lambda2\big)^2\psi
\;\Rightarrow\;
\Lambda_{\max}=\sqrt{-\mathcal E_0}-1 ,
\end{equation}
where $\mathcal E_0$ is the ground-state energy of the Schr\"odinger operator $-\partial_t^2-g^2C_{\phi'}(t)$. Hence, our generalized Schr\"odinger equation reduces to the two-replica result of SCS theory~\cite{Chaos-1988}.

\textit{Negative plasticity slows down chaotic fluctuations.}---
If $C_\phi$ varies slowly on the scale $\tau_s$ (correlation time $T_c\gg\tau_s$), then $K(u)\simeq\frac{\kappa}{\tau_s}e^{-u/\tau_s}$ with $\kappa=\lambda C_\phi(0)$, and $\tilde K(i\omega)=\frac{\kappa}{1+i\omega\tau_s}$. Hence,
\begin{equation}
\begin{split}
|\ell(i\omega)|^2&=\frac{(1-\tau_s\omega^2)^2+\omega^2(1+\tau_s-\kappa)^2}{1+\tau_s^2\omega^2}\\
&=1+\omega^2\big[(1-\kappa)^2-2\kappa\tau_s\big]+O(\omega^4). \label{eq:slowD}
\end{split}
\end{equation} 
The plastic loop acts as a second-order (mass--spring--friction) filter. Consequently, $|\ell(i\omega)|^2\simeq1+\tau_e^2\omega^2$ with $
\tau_e=\sqrt{(1+q|\lambda|)^2+2q|\lambda|\tau_s}\simeq q|\lambda|$, where $q\equiv C_{\phi}(0)$. This is the classical closure with the unit membrane time replaced by $\tau_e$. We thus have $q=q_{0}$ (the subscript means $\lambda=0$). According to the relaxation time definition $\tau_{\rm rel}=\int_0^\infty\left[C_\phi^{(0)}(t/\tau_e)/q_{0}\right]^2dt=\tau_e\tau_{\rm ref}^{(0)}(g)=\tau_{\rm rel}^{(0)}(g)q_{0}|\lambda|$. As $\lambda\to-\infty$, our slow dynamics assumption from the beginning is automatically satisfied. 

When $g\to1^+$ ($g-1=\epsilon\ll1$), $\tau_{\rm rel}^{(0)}$ also diverges as $\sqrt{3}/\epsilon$~\cite{Chaos-1988,PRE-2018}. Note that $q_0=\epsilon+O(\epsilon^2)$~\cite{PRE-2018,Zhang-2026}. We thus have plasticity-induced characteristic time $\tau_{\rm ref}(\lambda,g)=\sqrt{3}\sqrt{(|\lambda|+1/(g-1))^2+2\tau_s|\lambda|/(g-1)}$, which leads to the following intriguing results: (1) when $|\lambda|\ll1/(g-1)$, $\tau_{\rm ref}(\lambda,g)=\sqrt{3}/(g-1)+\sqrt{3}(1+\tau_s)|\lambda|$, implying that the critical slowing dominates and plasticity adds a correction linear in $|\lambda|$. (2) when $|\lambda|\gg1/(g-1)$, $\tau_{\rm ref}(\lambda,g)=\sqrt{3}|\lambda|+\sqrt{3}(1+\tau_s)/(g-1)$, implying that the plasticity-induced divergence dominates. The cross-over is set by $|\lambda|(g-1)\sim1$. 

\begin{figure}
\includegraphics[width=0.5\textwidth]{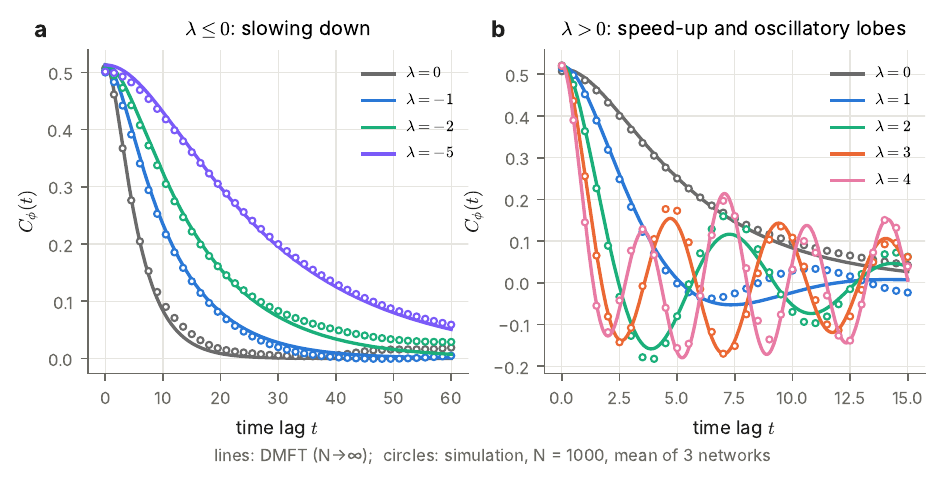}
\caption{\label{fig01}
Autocorrelation profile vs. time lag. Parameters: $(g,\tau_s)=(2.0,2.5)$. (a) negative plasticity. (b) positive plasticity. 
}
\end{figure}

\begin{figure*}
\includegraphics[width=\textwidth]{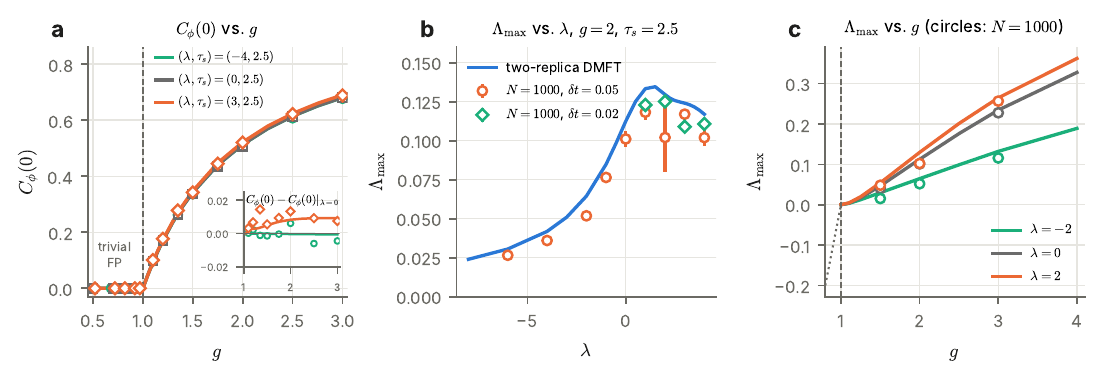}
\caption{\label{fig02}
Dynamics stability and maximum Lyapunov exponents. Symbols are finite-sized network simulations. (a) Activity variance vs. $g$. FP: fixed point. (b) Lyapunov exponents vs. $\lambda$. $\delta t$: discretization time step. (c) Lyapunov exponents vs. $g$.
}
\end{figure*}

\textit{Positive plasticity produces a preferred frequency.}---
As $\lambda$ increases, the activity spectrum $S_\phi(\omega)$ changes from a
single maximum at zero frequency to two symmetric maxima at
nonzero frequencies. We denote the spectral peak-splitting
threshold by $\lambda_{\mathrm{split}}(g)$.
Since $C_\phi$ is even,
$S_\phi(\omega)=2\int_0^\infty
C_\phi(\tau)\cos(\omega\tau)\,d\tau$, and hence $S_\phi''(0)=-2\int_0^\infty
    \tau^2 C_\lambda(\tau)\,d\tau$.
At fixed $g$, $\lambda_{\mathrm{split}}$ is the first value at
which this curvature changes from negative to positive as
$\lambda$ increases from zero, satisfying
\begin{equation}
    [S_\phi^{\lambda_{\mathrm{split}}}]''(0)=0
    \quad\Longleftrightarrow\quad
    \int_0^\infty\tau^2
    C_\phi^{\lambda_{\mathrm{split}}}(\tau)\,d\tau=0.
    \label{eq:splitting_condition}
\end{equation}

Furthermore, near the generic continuous splitting, the emerging peak frequency
and its height above the zero-frequency value obey $\omega_{\mathrm{peak}}
    \propto(\lambda-\lambda_{\mathrm{split}})^{1/2}$, and $S_\phi(\omega_{\mathrm{peak}})-S_\phi(0)
    \propto(\lambda-\lambda_{\mathrm{split}})^2$, as $\lambda\to\lambda_{\rm split}^+$. Technical details to derive this critical behavior are devoted to SM~\cite{SM}.

      \begin{figure}
\includegraphics[width=0.5\textwidth]{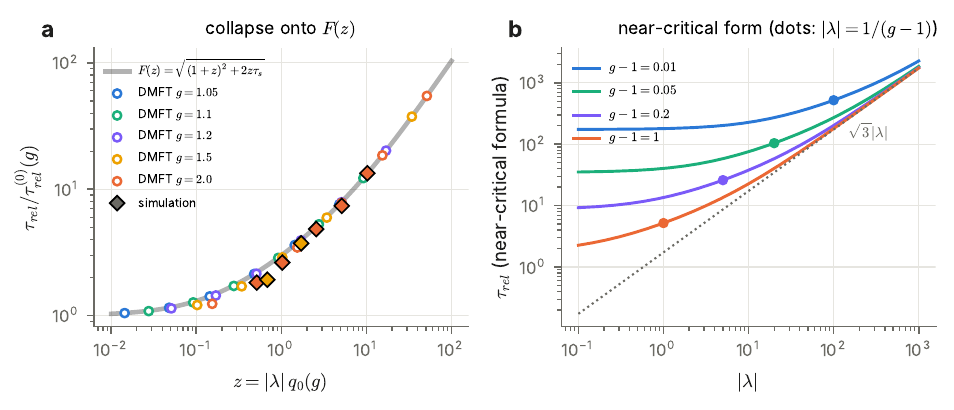}
\caption{\label{fig03}
Dynamics time scale tuned by negative plasticity.  (a) Collapse of rescaled characteristic time scale onto a universal $F(z)$ for different $g$.
Diamonds: simulations ($g=2$, $N=1000$; $g=1.5$, $N=2000$).
 (b) Near-critical time scale vs. $|\lambda|$.
}
\end{figure}

\textit{Results.}---
 We first show how the autocorrelation of neural activity is tuned by the plasticity sign and strength [Fig.~\ref{fig01}]. It is revealed that the negative plasticity slows down the chaotic fluctuation even at a large value of $g$. Therefore, in a learning system, the edge of chaos may not be a uniquely optimal regime because the relaxation time can also diverge far away from the edge. On the other side, the positive plasticity prefers an oscillatory behavior for single neurons. DMFT theory predicts the simulation on finite-sized networks very well. 

  We next study the transition point. As revealed in our theory, the stability of the trivial state (including neural and synaptic states) is the same as in the classic non-learning system, i.e., $g_c=1$ [see Fig.~\ref{fig02} (a)], which coincides with the chaos transition predicted by the generalized Schr\"odinger equation (see Fig.~\ref{fig02} (c)]. Remarkably, chaos is weakest for $\lambda\to-\infty$ and strongest near $\lambda\simeq1.5$, which is followed by a slow decay [Fig.~\ref{fig02} (b)]. 
  
     \begin{figure*}
\includegraphics[width=\textwidth]{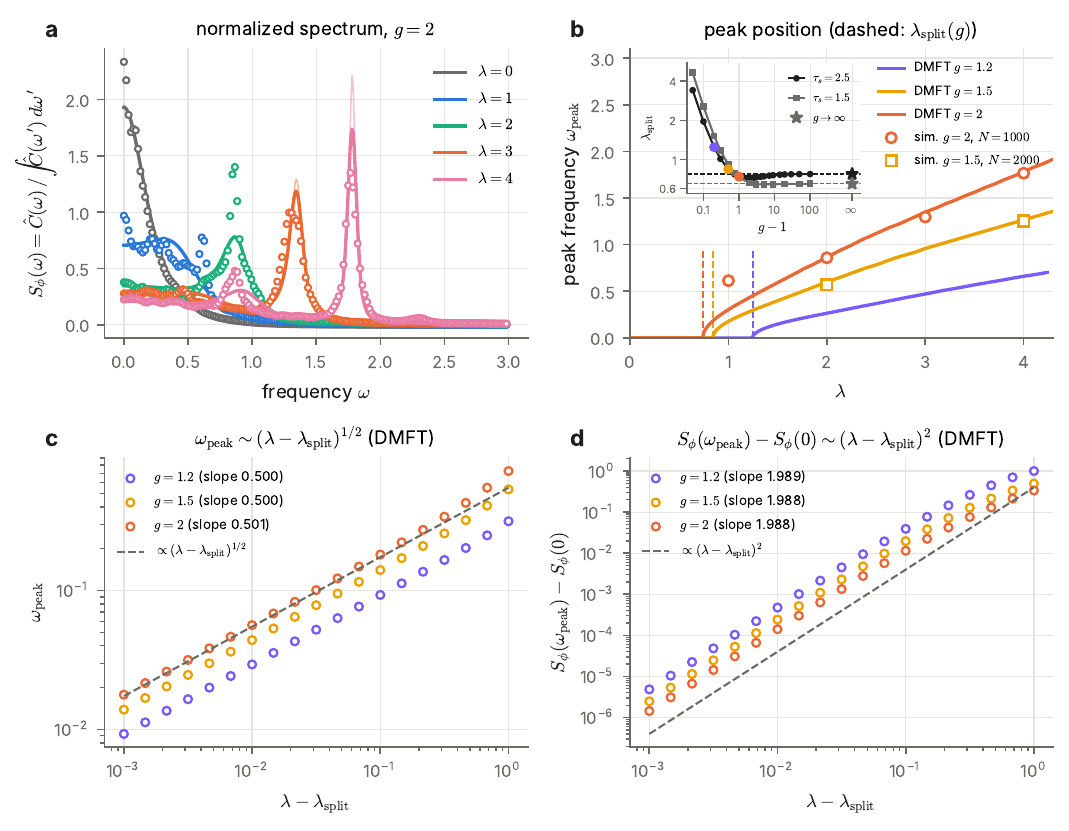}
\caption{\label{fig04}
Normalized power spectrum at positive plasticity. (a) DMFT compared with simulations (circles, $N=1000$). (b) Peak frequency vs. $\lambda$. The inset shows 
$\lambda_{\rm split}(g)$. (c) Scaling behavior of $\omega_{\rm peak}$. (d) Scaling behavior of $S_\phi(\omega_{\rm peak})-S_\phi(0)$.
}
\end{figure*}

 The theory predicts a collapse of the exact DMFT relaxation time scale [$\tau_{\rm rel}(\lambda,g)/\tau_{\rm rel}^{(0)}(g)$] onto a universal function $F(z)=\sqrt{(1+z)^2+2z\tau_s}$, supported by finite-sized network simulation results (see Fig.~\ref{fig03}). In the classic non-learning scenario, the relaxation time diverges near $g_c$ as well. Our theory predicts a crossover from critical to plastic slowing down at $|\lambda|=1/(g-1)$, which is shown in Fig.~\ref{fig03}.
 
 Finally, we study the power spectrum at positive plasticity (Fig.~\ref{fig04}). Positive plasticity promotes a preferred frequency. The oscillation is a property of single neurons, not a 
 population collective rhythm. The single-neuron spectra concentrate at the predicted frequency [see Fig.~\ref{fig04} (b)] (data shown in SM~\cite{SM}). Near $g_c$, the correlation time diverges. When $\tau_e^2=(1-\kappa)^2-2\kappa\tau_s$ changes sign from positive to negative, the power spectra would develop a maximum away from $\omega=0$. Thus $\lambda_{\rm split}(g)\simeq\frac{\kappa_p}{q(g)}$, where $\kappa_p$ is the smaller root of $\tau_e^2=0$, and $q(g)=C_\phi(0)\propto g-1$ near $g_c$~\cite{Zhang-2026}. This divergent behavior is predicted in the inset of Fig.~\ref{fig04} (b). $\lambda_{\rm split}$ falls steeply from the critical divergence and reaches a shallow minimum near $g\approx 2.8$ and finally approaches $\lambda_{\rm split}(\infty) =0.774$ as $g\to\infty$. The minimum marks the crossover from amplitude-dominated ($C_\phi(0)$ saturates) to shape-dominated (steepening of $\phi$) behavior. The crossover occurs at a larger value of $g$ for smaller $\tau_s=1.5$. Figure~\ref{fig04} shows the critical behavior $\omega_{\rm peak}$ and power spectrum near to $\lambda_{\rm split}$ with critical exponents $1/2$ and $2$, respectively.

\textit{Concluding remarks.}---
In this work, we develop a generalized dynamical mean-field theory of learning in plastic recurrent neural networks. The synaptic plasticity is neural-speed driven and not symmetric as in Hebbian learning. The neural state is thus tuned by the velocity feedback. For negative plasticity, this feedback is friction: the dynamics slow down as if time is dilated. Both $\tau_{\rm rel}$ and $1/\Lambda_{\max}$ diverge as $\lambda\to-\infty$. For positive plasticity, this feedback is inertia: the filter $|\ell(i\omega)|^{-2}$ becomes resonant, reflected in the activity power spectrum, and single neurons oscillate at $\omega_{\rm peak}$. The chaos transition point and the amplitude of the chaotic state near $g_c$ are all independent of $\lambda$ and $\tau_s$. However, chaos is weakest for $\lambda\to-\infty$ and strongest at some intermediate positive $\lambda$. Our theory therefore establishes a non-equilibrium statistical mechanics framework to treat complex neural dynamics coupled with asymmetric plasticity rules.

\section{Acknowledgments}
This research was supported by the National Natural Science Foundation of China for
Grant number 12475045, and Guangdong Provincial Key Laboratory of Magnetoelectric Physics and Devices (No. 2026B1212170049). 

\section{AI use statement}
The results reported in this study were generated through extensive interaction with the powerful AI models GPT-6.1 sol and Claude Opus 5.5. These models were used for theoretical derivations and coding, guided by the authors. The authors take full responsibility for the scientific content. This manuscript reports our own thoughts on the problem defined by the senior author.




\begin{thebibliography}{10}

\bibitem{Vyas-2020}
Saurabh Vyas, Matthew~D. Golub, David Sussillo, and Krishna~V. Shenoy.
\newblock Computation through neural population dynamics.
\newblock {\em Annual Review of Neuroscience}, 43:249--275, 2020.

\bibitem{Ostojic-2024}
Srdjan Ostojic and Stefano Fusi.
\newblock Computational role of structure in neural activity and connectivity.
\newblock {\em Trends in Cognitive Sciences}, 28:677, 2024.

\bibitem{Huang-2024}
Haiping Huang.
\newblock Eight challenges in developing theory of intelligence.
\newblock {\em Front. Comput. Neurosci}, 18:1388166, 2024.

\bibitem{DWK-2025}
Wenkang Du and Haiping Huang.
\newblock Response function as a quantitative measure of consciousness in brain
  dynamics.
\newblock {\em Phys. Rev. Res.}, 7:043249, Dec 2025.

\bibitem{Abbott-2009}
David Sussillo and Larry~F Abbott.
\newblock Generating coherent patterns of activity from chaotic neural
  networks.
\newblock {\em Neuron}, 63(4):544--557, 2009.

\bibitem{JPA-1993}
R~W Penney, A~C~C Coolen, and D~Sherrington.
\newblock Coupled dynamics of fast spins and slow interactions in neural
  networks and spin systems.
\newblock {\em Journal of Physics A: Mathematical and General}, 26(15):3681,
  1993.

\bibitem{JPA-1994}
R~Penney and D~Sherrington.
\newblock Slow interaction dynamics in spin-glass models.
\newblock {\em Journal of Physics A: Mathematical and General}, 27(12):4027,
  1994.

\bibitem{Clark-2024}
David~G. Clark and L.~F. Abbott.
\newblock Theory of coupled neuronal-synaptic dynamics.
\newblock {\em Phys. Rev. X}, 14:021001, 2024.

\bibitem{Jstat-2023}
Samantha~J Fournier and Pierfrancesco Urbani.
\newblock Statistical physics of learning in high-dimensional chaotic systems.
\newblock {\em Journal of Statistical Mechanics: Theory and Experiment},
  2023(11):113301, 2023.

\bibitem{Du-2024}
Wenkang Du and Haiping Huang.
\newblock Synaptic plasticity alters the nature of the chaos transition in
  neural networks.
\newblock {\em Phys. Rev. E}, 112:054208, 2025.

\bibitem{Qiu-2024}
Junbin Qiu and Haiping Huang.
\newblock An optimization-based equilibrium measure describing fixed points of
  non-equilibrium dynamics: application to the edge of chaos.
\newblock {\em Commun. Theor. Phys.}, 77(3):035601, 2025.

\bibitem{Huang-2025}
Weizhong Huang and Haiping Huang.
\newblock Freezing chaos without synaptic plasticity.
\newblock {\em Phys. Rev. E}, 112:044227, Oct 2025.

\bibitem{Zhang-2026}
Li-Ru Zhang and Haiping Huang.
\newblock Kinetic energy in random recurrent neural networks.
\newblock {\em Phys. Rev. E}, 114:034207, 2026.

\bibitem{Yu-2025}
Zhendong Yu, Weizhong Huang, and Haiping Huang.
\newblock Neural langevin machine: a local asymmetric learning rule can be
  creative.
\newblock {\em arXiv:2506.23546}, 2025.

\bibitem{Senn-2014}
Robert Urbanczik and Walter Senn.
\newblock Learning by the dendritic prediction of somatic spiking.
\newblock {\em Neuron}, 81(3):521--528, 2014.

\bibitem{Senn-2024}
Walter Senn, Dominik Dold, Akos~F Kungl, Benjamin Ellenberger, Jakob Jordan,
  Yoshua Bengio, João Sacramento, and Mihai~A Petrovici.
\newblock A neuronal least-action principle for real-time learning in cortical
  circuits.
\newblock {\em eLife}, 12:RP89674, 2024.

\bibitem{Chaos-1988}
H.~Sompolinsky, A.~Crisanti, and H.~J. Sommers.
\newblock Chaos in random neural networks.
\newblock {\em Phys. Rev. Lett.}, 61:259--262, 1988.

\bibitem{SM}
See the supplemental material at http://... for technical details and
  simulation methods.

\bibitem{Zou-2024}
Wenxuan Zou and Haiping Huang.
\newblock {Introduction to dynamical mean-field theory of randomly connected
  neural networks with bidirectionally correlated couplings}.
\newblock {\em SciPost Phys. Lect. Notes}, page~79, 2024.

\bibitem{MSR-1973}
P.~C. Martin, E.~D. Siggia, and H.~A. Rose.
\newblock Statistical dynamics of classical systems.
\newblock {\em Phys. Rev. A}, 8:423--437, 1973.

\bibitem{Sommers-1987}
Hans-J\"urgen Sommers.
\newblock Path-integral approach to ising spin-glass dynamics.
\newblock {\em Phys. Rev. Lett.}, 58:1268--1271, 1987.

\bibitem{Jans-1976}
Hans-Karl Janssen.
\newblock On a lagrangean for classical field dynamics and renormalization
  group calculations of dynamical critical properties.
\newblock {\em Zeitschrift f{\"u}r Physik B Condensed Matter}, 23(4):377--380,
  1976.

\bibitem{Domin-1978}
C.~De~Dominicis.
\newblock Dynamics as a substitute for replicas in systems with quenched random
  impurities.
\newblock {\em Phys. Rev. B}, 18:4913--4919, 1978.

\bibitem{Yu-2025b}
Zhendong Yu and Haiping Huang.
\newblock Network reconstruction may not mean dynamics prediction.
\newblock {\em Physical Review E}, 111(3):034308, 2025.

\bibitem{Huang-2022}
Haiping Huang.
\newblock {\em Statistical Mechanics of Neural Networks}.
\newblock Springer, Singapore, 2022.

\bibitem{PRE-2018}
A.~Crisanti and H.~Sompolinsky.
\newblock Path integral approach to random neural networks.
\newblock {\em Phys. Rev. E}, 98:062120, 2018.

\end{thebibliography}


\end{document}